\documentclass[12pt]{article}

\usepackage{newtxtext,newtxmath}

\usepackage{graphicx}
\usepackage[T1]{fontenc} 

\usepackage[letterpaper,margin=1in]{geometry}

\renewenvironment{abstract}
	{\quotation}
	{\endquotation}

\date{}

\makeatletter
\renewcommand{\fnum@figure}{\textbf{Figure \thefigure}}
\renewcommand{\fnum@table}{\textbf{Table \thetable}}
\makeatother

\usepackage{scicite}

\usepackage{url}

\def\scititle{
	Wafer-Scale Single-Crystalline Monolayer Graphene
}
\title{\bfseries \boldmath \scititle}

\author{
	Johanna~Huhtasaari$^{1}$, 
	Joyal~Jain Palakulam$^{1}$,  
    Awse Salha$^{1}$,  \and
    Per Hyldgaard$^{1}$,
    Elsebeth Schröder$^{1}$,  
	Magnus~Hårdensson~Berntsen$^{2}$, \and
        Oscar~Tjernberg$^{2}$, 
        Manasi~Shah$^{3}$,
        Rodrigo~Martinez-Duarte$^{3}$,\and
        Hans He$^{4}$, 
        Johannes Hofmann$^{5,6}$,
        Thilo Bauch$^{1}$,
        Naveen~Shetty$^{1}$,\and
        Samuel~Lara-Avila$^{1,\ast}$ \and
	\small$^{1}$Microtechnology and Nanoscience, Chalmers University of Technology, 412 96 Gothenburg, \ Sweden\and
        \small$^{2}$Department of Applied Physics, KTH Royal Institute of Technology, 114 19 Stockholm, Sweden\and
        \small$^{3}$ Department of Mechanical Engineering, Clemson University, Clemson, USA\and
    \small$^{4}$RISE (Research Institutes of Sweden), \ 501 15 Borås, Sweden \and
    \small$^{5}$Department of Physics, Gothenburg University, 412 96 Gothenburg, Sweden\and
        \small$^{6}$Nordita, Stockholm University and KTH Royal Institute of Technology, 106 91 Stockholm, Sweden\and
	\small$^\ast$Corresponding author. Email: samuel.lara@chalmers.se \and
}

\begin{document} 

\maketitle

\begin{abstract} \bfseries \boldmath


Producing large-area single-crystalline graphene is key to realizing its full potential in advanced applications, including twistronics. Yet, controlling graphene growth kinetics to avoid grain boundaries or multilayer growth remains challenging. Here, we demonstrate single-crystalline graphene free from multilayer domains via one-step delamination of epitaxial graphene from silicon carbide (SiC). This is enabled by a specific surface reconstruction of 4H-SiC(0001) achieved in our growth conditions. High crystalline quality is confirmed by the observation of the half-integer quantum Hall effect -- the hallmark of monolayer graphene -- in near cm-sized crystals. The scalability of our process, explored with 4''-wafers, represents an advance toward large-scale integration of high-performance graphene applications.


\end{abstract}

\noindent

Realizing scalable applications of two-dimensional (2D) materials in electronics, photonics, and quantum technologies requires large-area single-crystalline films \cite{Zhao2024GrapheneBenches}. In the atomically-thin limit, grain boundaries strongly affect 2D material properties: they lower carrier mobility (detrimental for high-performance electronics) \cite{Yu2011ControlDeposition,Zhang2014GrainGraphene,Ma2017DirectGraphene} and mechanical strength \cite{Zhang2012IntrinsicBoundaries}, increase chemical reactivity (compromising robustness to harsh processing conditions) \cite{King2008ObservationsPolycrystal,Yasaei2014ChemicalBoundaries}, and introduce material property variability that complicates reproducible device manufacturing \cite{Canto2025Multi-projectProject}. Significant progress in synthesizing large-area single-crystals, particularly graphene, has been achieved through chemical vapor deposition (CVD) \cite{Wang2021Single-crystalGraphene, Amontree2024ReproducibleDeposition, Xue2025DirectSubstrate}. A driving force behind this progress is the versatility of the grown material in being transferable to virtually any substrate. Yet, key challenges in scalable CVD growth of single-crystalline graphene include an extensive parameter space for growth optimization (e.g. controlling nucleation of carbon atoms and domain growth), and the requirement to develop and use expensive single-crystalline metal catalysts to template the assembly of carbon atoms \cite{Xue2025DirectSubstrate,Zhang2020ControlledFilms}. In contrast, epitaxial graphene (epigraphene) growth on silicon carbide (SiC) capitalizes on the commercial availability of single-crystalline SiC substrates, resulting in large-area single-crystalline graphene layers \cite{Norimatsu2014EpitaxialPerspectives,Yazdi2016EpitaxialCharacterization}. Yet, the delamination and transfer of graphene from SiC remains less explored. So far, successful attempts have faced two grand challenges: (i) material degradation due to the harsh delamination and transfer process results in defected graphene with very low carrier mobilities \cite{Unarunotai2009TransferTransistors,Unarunotai2010Layer-by-layerWafer}, and (ii) multilayer graphene domains (i.e. graphene patches) -- a ubiquitious defect in epigraphene -- co-transfer with the top graphene layer\cite{Kim2013Layer-ResolvedLayers,Bae2017UnveilingGraphene}. The presence of excess graphene layers, cumbersome to control also in CVD growth, causes electronic inhomogeneity that limits high-performance applications \cite{Yager2013ExpressTransport}. Furthermore, for future twisted-layer devices, avoiding grain boundaries and multilayer patches is essential \cite{He2021MoireReview}. 

Here, we demonstrate the possibility to produce patch-free large-area graphene single-crystals by selectively delaminating the topmost graphene layer grown on SiC, while preserving its high electronic quality. Importantly, this relaxes the need for complex graphene growth optimization -- via epitaxial or CVD methods -- aiming to suppress multilayer growth \cite{Kruskopf2016ComebackSiC} and grain boundaries. We verify the quality of the transferred monolayer using a wide array of techniques, including electrical transport measurements. Remarkably, we observe the half-integer quantum Hall effect in centimeter scale samples. This confirms that the high electronic quality and single-crystallinity of the transferred graphene is preserved, because the presence of grain boundaries or excessive disorder would prevent exact Hall quantization \cite{Cummings2014QuantumBoundaries, Bergvall2015InfluenceGraphene}.  

\subsection*{Transfer of single-crystalline monolayer graphene.}
We find that the success in selective one-step delamination and transfer of monolayer graphene from the substrate relies on a specific surface reconstruction -- hereafter called a Type A surface --,  achieved on the silicon-terminated face of 4H-SiC in our growth conditions \cite{Virojanadara2008Homogeneous6H-SiC0001} (see Methods). 
During growth, the SiC surface reconstructs, followed by sublimation of silicon atoms (the so-called step-bunching stage), and then the remaining carbon atoms crystallize into a buffer layer with the continuous monolayer graphene atop  \cite{Norimatsu2014EpitaxialPerspectives,Yazdi2016EpitaxialCharacterization}. Our growth process minimizes step bunching and nearly eliminates multilayer formation. We identify these as hallmarks of the Type A surface, which we obtain at temperatures above $T_0=1750^\circ$C. The Type A surface is key for achieving selective monolayer delamination, and the process is robust even if we deviate from $T_0$, allowing to significantly relax epigraphene growth temperature optimization. Only large temperature excursions ($T \gtrsim
 T_0+150$$^\circ$C) cause another distinct surface reconstruction -- hereafter called Type B -- that reduces the selectivity of the delamination process. The Type A surface is highly reproducible for 7 mm $\times$ 7 mm-sized chips in our growth conditions  \cite{Shetty2024UltralowDevices}. We have observed selective delamination in over 25 chips, and here we report detailed characterization results from 10 of these.

\begin{figure} 
	\centering
	\includegraphics{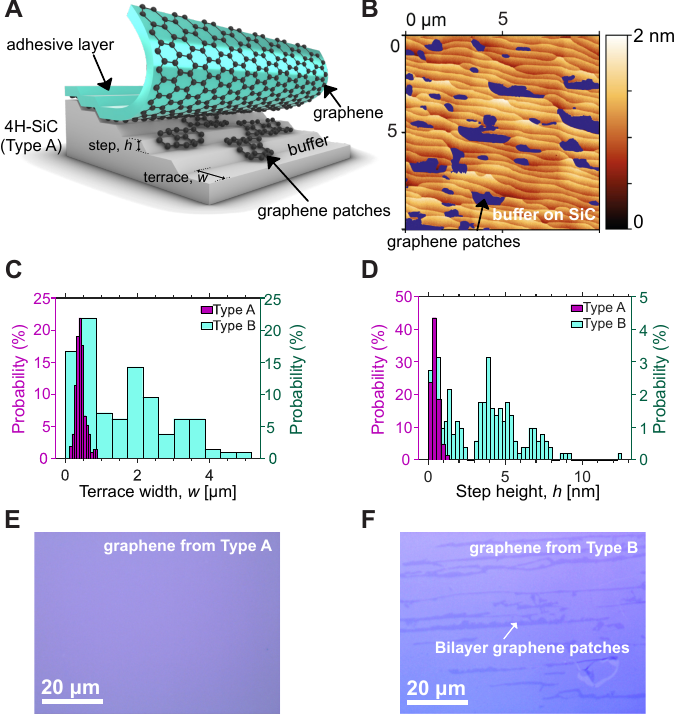} 
	\caption{\textbf{Production of single-crystalline monolayer graphene.} (\textbf{A}) Illustration of delamination of epigraphene from a Type A SiC surface, where excess graphene patches are left behind on the substrate, i.e. on the buffer layer. (\textbf{B}) AFM topography map of the Type A SiC-substrate after graphene delamination, where all the excess graphene patches remain after delamination; these are overlaid in false-color (blue). (\textbf{C}) Terrace width histogram obtained from AFM data, normalized to probability, for Type A and B SiC surface reconstructions.  (\textbf{D}) Same as (C) for step height. (\textbf{E}) Optical micrograph of  epigraphene delaminated from a Type A SiC substrate and transferred to SiO$_2$/Si. Note the absence of graphene patches. (\textbf{F}) Same as (E) with graphene delaminated from a Type B surface, where bilayer patches co-delaminated with the top graphene layer. 
    }
	\label{fig:1} 
\end{figure}

Figure \ref{fig:1}A illustrates the delamination principle, where epigraphene is covered with an "adhesive layer": a thin metal film ($t>20$ nm) directly deposited on graphene, with a thermal-release tape for mechanical support (see Methods). Notably, AFM inspection (Fig. \ref{fig:1}B) reveals that all excess graphene patches (false-colored) remain on the SiC substrate after delaminating the topmost graphene layer from a Type A surface (for a full set of AFM images, see Fig. \ref{subfig:1}). To quantitatively characterize the morphology of the Type A surface, Figure \ref{fig:1}D shows histograms of terrace width for Type A epigraphene grown at relaxed temperature conditions, $T_0+50$$^\circ$C, compared to the surface of Type B epigraphene grown at a higher temperature $T_0+150$$^\circ$C. Type A surfaces exhibit narrow terraces ($w \lesssim 1$ $\mu$m), whereas the Type B surfaces exhibit wider terraces with a broader distribution up to $w \lesssim 5$ $\mu$m. The corresponding step height histograms (Fig. \ref{fig:1}D) reveal shallow steps for the Type A surface, with $\sim75\%$ of the steps heights being half a 4H-SiC unit cell ($\sim 0.5$ nm). Type B exhibits higher steps, exceeding 12 4H-SiC-unit cells ($\sim 12$ nm). Altogether, the Type A surface morphology allows for growth of narrow graphene patches. Indeed, extensive AFM characterization of Type A SiC surface after graphene delamination reveals that all of the patches found display a width $w < 1.4$ $\mu$m (Fig. \ref{SI_fig_histogram_patchsize_typeA}). 

The stark difference between Type A and B surfaces in terms of selective delamination is evident from optical microscope inspection of the transferred graphene layer. Figure \ref{fig:1}E shows a representative graphene monolayer transferred onto SiO$_2$/Si. This monolayer, derived from a Type A surface, is entirely free of multilayer domains. In contrast, Figure \ref{fig:1}F shows a representative graphene layer transferred from a Type B surface, with excess graphene patches appearing as horizontal dark stripes. These patches co-delaminated with the top graphene layer, an issue reported in earlier work \cite{Kim2013Layer-ResolvedLayers,Bae2017UnveilingGraphene}.


\subsection*{Surface characterization of transferred graphene.}

In sharp contrast to earlier reports of graphene delamination from SiC \cite{Kim2013Layer-ResolvedLayers} and graphite \cite{Moon2020Layer-engineeredGrapheneb}, we find that Type A surfaces enable several metals to selectively delaminate monolayer graphene, with gold in particular resulting in excellent crystalline quality. Figure \ref{fig:2}A shows a comparison of Raman spectra collected for graphene transferred onto SiO$_2$/Si using gold, silver, and nickel. For reference, the uppermost panel in Figure \ref{fig:2}A shows the characteristic spectral features of epigraphene, namely, a broadened and suppressed 2D-peak \cite{Rohrl2008RamanSiC0001,Ni2008RamanSubstrate,Lee2008RamanSiO2}, and buffer layer bands in the 1350–1570 cm$^{-1}$ range \cite{Fromm2013ContributionSiC0001, Milenov2024RamanTheory}. After gold-assisted transfer to SiO$_2$/Si, the typical Raman features are similar to those of graphene exfoliated from high-quality graphite crystals \cite{Ferrari2013RamanGraphene}, with high crystalline quality indicated by absence of the defect-induced D-peak ($\sim 1350$ cm$^{-1}$), and a sharp, symmetric 2D-peak with full-width at half-maximum (FWHM) $28$ cm$^{-1}$. Figure \ref{fig:2}A further shows that silver and nickel, which also enable highly selective monolayer delamination (later demonstrated in Figs. \ref{fig:3}H and \ref{SI:AFM_Ni}), result in defected graphene, as indicated by a noticeable D-peak (most pronounced in the case of nickel).
Additionally, we note that all graphene Raman peaks red-shift after transferring graphene from SiC to SiO$_2$/Si, likely due to the release of compressive strain upon delamination from the SiC substrate \cite{Rohrl2008RamanSiC0001,Lee2008RamanSiO2}. A more detailed Raman analysis \cite{Lee2012OpticalGraphene} (see Fig. \ref{supplementary:Raman_strain_doping_analysis_PLOT}) indicate compressive strain for epigraphene, and tensile strain and p-doping for graphene on SiO$_2$/Si. 
Figure. \ref{fig:2}B shows a selected-area electron diffraction (SAED) pattern of graphene delaminated from SiC transferred onto an amorphous carbon-TEM grid, showing diffraction spots of graphene with lattice parameter $a = 2.54\pm 0.02$ Å (see Methods), higher than the theoretical value ($a = 2.46$ Å), indicating a graphene lattice under tensile strain. 

The linear Dirac dispersion of the transferred graphene is confirmed by ARPES measurements performed with $E=18.1$ eV photons and spot size $\sim 100$ $\mu$m \cite{Guo2022ASpectroscopy}. Figure \ref{fig:2}C shows the typical band-structure of as-grown epigraphene, with Fermi level ($E_F$) positioned $400$ meV above the Dirac point, consistent with other reports on epigraphene \cite{Starke2009EpitaxialElectronics}, corresponding to an electron density $n =\frac{1}{\pi}\left(\frac{E_F}{\hbar v_F}\right)^2\approx 7.0 \times 10^{12}$ cm$^{-2}$ \cite{CastroNeto2009TheGraphene} with the Fermi velocity $v_F = 1.3\times10^6$ ms$^{-1}$ extracted from the fit of the ARPES peak intensity positions (see Supplementary Text). After Au-assisted transfer onto SiO$_2$/Si (Fig. \ref{fig:2}D), graphene turns p-doped due to interaction with the oxide \cite{Nistor2012DopingSubstrates}. Interestingly, the Fermi velocity of carriers nearly doubles to $v_F = 2.2\times10^6$ ms$^{-1}$, approaching values reported for suspended graphene \cite{Elias2011DiracGraphene}. This increase, along with the momentum broadening $\Delta k_{\parallel}=0.11$ Å$^{-1}$, small for graphene on a SiO$_2$/Si substrate, is a consequence of high crystallinity and reduced electron-electron screening in the local dielectric environment \cite{Hwang2012FermiModification}. We note that the suppressed photoemission intensity of the Dirac state at positive momenta in Figure \ref{fig:2}D arises from the experimental geometry, and the right-branch dispersion remains traceable at higher binding energies (see Fig. \ref{SI:ARPES_fig}). Altogether, the combined electron diffraction, Raman, and ARPES analyses demonstrate that the crystalline quality of graphene is retained after delamination and transfer from SiC substrate.

\subsection*{Electrical transport of transferred graphene.}

Magneto-transport measurements reveal that epigraphene transferred onto SiO$_2$/Si exhibits transport properties that are undistinguishable from those of high-quality exfoliated graphene flakes on the same substrate. Figure \ref{fig:3}A shows a typical Hall bar device on SiO$_2$/Si produced by gold-assisted delamination. To protect graphene from the ambient, devices were passivated with PMMA resist before measurements of the longitudinal resistivity, $\rho_{xx}$ ($=R_{xx}W/L$), and transverse resistance, $R_{xy}$. Fits of $\rho_{xx}$ to the field-effect response \cite{Gosling2021UniversalImpurities} in Figure \ref{fig:3}B (Eq. \ref{eq:S4}) yield a carrier mobility $\mu_{FET} =8,670$ cm$^2$V$^{-1}$s$^{-1}$ at 300 K, in agreement with those extracted from low-field ($|B| <1$ T) Hall effect measurements. The Hall mobilities ($\mu_H = \vert R_H \vert/\rho_{xx}(B=0\,\mathrm{T})$) exhibit only a weak linear decrease with carrier density ($n_H=\frac{1}{e R_H}$) in the range $n_H \approx 1$–$4\times 10^{12}$ cm$^{-2}$, as shown in Figure \ref{fig:3}C. The carrier mobility is limited by charged impurity scattering, typical of graphene on SiO$_2$/Si \cite{Adam2007ATransport, Chen2008Charged-impurityGraphene}, with a charged impurity density $n_{\mathrm{imp}} \sim 5-6 \times 10^{11}$ cm$^{-2}$ obtained from fits to $\mu_{FET}$ and $\mu_H$ (see Tables S1-S2). The values of room-temperature carrier mobilities are high considering substrate-scattering effects, metal residues (see Fig. \ref{supplementary_fig:AFM_transferred}) and the polymer encapsulating layer. At low temperatures, these limit the mobility values to $\mu_{FET} = 11,000$ cm$^2$V$^{-1}$s$^{-1}$ and $\mu_H= 12,500$ cm$^2$V$^{-1}$s$^{-1}$ at $T=2$ K. Altogether, the observed room-temperature carrier mobilities in transferred graphene approach the upper limit for graphene on SiO$_2$/Si substrates \cite{Chen2008IntrinsicSiO2,Tyagi2022Ultra-cleanSubstrates}, and exceed typical mobilities reported for epigraphene, which are in the 1,000 cm$^2$V$^{-1}$s$^{-1}$-range \cite{Norimatsu2023ACarbide}.

The observation of the half-integer quantum Hall effect (QHE) gives unequivocal evidence that all electronic properties of graphene are preserved after transfer from SiC  \cite{Novoselov2005Two-dimensionalGraphene,Zhang2005ExperimentalGraphene}. Figure \ref{fig:3}D presents the gate response of $\rho_{xx}$ and $R_{xy}$, measured at $T=2$ K and $B=14$ T, which shows quantum Hall plateaus in $R_{xy}$ and vanishing $\rho_{xx}$ at filling factors $|\nu| = 2, 6, 10 $ ($\nu = 4N + 2$ with Landau index $N$) \cite{Zhang2005ExperimentalGraphene}. The observation of QH plateaus for filling factors $\vert \nu \vert >2$ is noteworthy because they are difficult to observe for epigraphene due to disorder (intervalley scattering) caused by the stepped SiC substrate \cite{Lara-Avila2011DisorderedMeasurements}. This indicates that the corrugation of epigraphene due to the steps in SiC and the concomitant interaction with the buffer layer \cite{DeJong2023StackingCarbide} is not inherited by the transferred material (this is also supported by AFM data in Fig. \ref{supplementary_fig:AFM_transferred}).

Higher-order Landau levels up to filling factor $\nu = 74$ are also observable in the magneto-resistance of the transferred graphene device. Figure \ref{fig:3}E shows Shubnikov-de Haas (SdH)-oscillations for electrons in the metallic limit ($n_H \approx 3.8\times10^{12}$ cm$^{-2}$) at different temperatures, persisting up to $T=150$ K. At large $N$, Landau levels are closely spaced in energy, so their manifestation as quantum oscillations in $\rho_{xx}$ require a small disorder-induced energy broadening. To quantify the energy broadening, we analyzed the temperature- and field-dependence of the oscillation amplitudes within the Lifshitz-Kosevich formalism \cite{Ando1982ElectronicSystems,Tan2011Shubnikov-deBias} (see Methods and Fig. \ref{SI_sdH_temp_dependence_fig}). From this, we estimate the cyclotron mass to $m_c/m_e=0.044 \pm 0.0056$ with $m_e$ the free-electron mass, lifetime $\tau = 50.8 \pm 6.59$ fs, and Dingle temperature $T_D = 23.9 \pm 3.10$ K, with the errors given as the expanded uncertainty with coverage factor $k=1$. These are typical values for exfoliated graphene on SiO$_2$/Si at similar carrier densities \cite{Novoselov2005Two-dimensionalGraphene,Tan2011Shubnikov-deBias}. Figure \ref{fig:3}F shows a plot of $\frac{1}{B_N}$ versus $N$, where $B_N$ denotes the magnetic field strength corresponding to Landau index $N$ at an SdH-oscillation minima in $\rho_{xx}$ at $T = 2$ K (see full data in Fig. \ref{subfig:sdH_small_device_holes}). A linear fit yields carrier densities $n_s$ (Eq. \ref{ns}) in agreement with those from linear Hall slopes ($n_H$). Extrapolating to $1/B_N = 0$ yields the intercept $N \approx -\tfrac{1}{2}$ reflecting the Berry phase of $\pi$ in monolayer graphene \cite{Novoselov2005Two-dimensionalGraphene,Zhang2005ExperimentalGraphene}. 

We use quantum transport to verify the single-crystallinity of our large-area graphene, through the demonstration of the half-integer quantum Hall effect in nearly centimeter-sized devices. A zero-resistance state ($\rho_{xx} = 0$) has been difficult to achieve in graphene produced by any other method, notably CVD graphene, due to grain boundaries \cite{Lafont2014AnomalousDeposition, Cummings2014QuantumBoundaries,Chau2022QuantumBoundary,Shen2011QuantumFilmsb}, or even epigraphene due to bilayer patches \cite{Bergvall2015InfluenceGraphene,Yager2013ExpressTransport}. Yet, graphene transferred from SiC allows for $\rho_{xx}=0$ even when silver, known to give slightly defective graphene (see Fig. \ref{fig:2}A), is used as the delamination layer. Figure \ref{fig:3}G shows a 7 mm $\times$ 7 mm SiO$_2$/Si substrate covered with silver-assisted transferred graphene, fitted with electrical contacts by shadow mask deposition. Room-temperature mobilities extracted from van der Pauw-measurements \cite{Rietveld2003DCGeometry} are of the order of 1,000 cm$^2$V$^{-1}$s$^{-1}$  (see Fig. \ref{supplementary_fig:largemobility_RvsVg}). At low temperatures, the gate voltage response of $\rho_{xx}$ and $R_{xy}$ (Fig. \ref{fig:3}H) exhibits QH plateaus for filling factors $\vert \nu\vert = 2, 6, 10, 14$. We verified the zero-resistance state by measuring the current-voltage (I-V) curves at the $\nu = \pm 2$-plateaus across diagonal contacts $V_{xy}^+$ and $V_{xy}^-$. Figure \ref{fig:3}I shows the I-V-curve in logarithmic scale (see plot in linear scale in Fig. \ref{subfig:IV})  that confirms the non-dissipative state for both holes and electrons, limited by the noise floor of $V_{rms}=83 $ nV. The voltage signal increases exponentially for currents above a critical value ($I_c\approx 3$ $\mu$A), indicating breakdown of the QHE. Significantly, the zero-resistance state provides evidence for the absence of grain boundaries or major doping or structural inhomogeneities across the entire graphene area, because these would short QH edge currents and result in a finite longitudinal resistance \cite{Lafont2014AnomalousDeposition,Cummings2014QuantumBoundaries}.

\subsection*{Wafer-scale graphene transfer.}
We demonstrate the scalability of our process through the delamination and transfer of epigraphene grown on recycled 4''-SiC wafers (see Methods), shown in Figure \ref{fig:4}A. Epigraphene growth on recycled wafers results in a surface morphology where up to 40$\%$ of the area contains excess graphene (see Fig. \ref{fig:4}B). The transferred graphene, in contrast, exhibits a remarkably low bilayer content. We quantified this by optical mapping of graphene transferred onto SiO$_2$/Si substrate (Fig. \ref{fig:4}C) via the optical contrast difference between mono- and bilayer graphene \cite{Yager2013ExpressTransport}. In total, we sampled 300 microscope images on the 4"-wafer, each corresponding to an area of 290 $\times$ 190 $\mu$m$^2$. An example with 1$\%$ bilayer content is shown in Figure \ref{fig:4}D, and the result of the wafer mapping is shown in Figures \ref{fig:4}E (color map) and  \ref{SI_fig_histogram_Bilayer_content_wafer_mapping} (histograms). The median value of the bilayer graphene content of the transferred graphene is $4.3\%$, implying a $\sim$10-fold reduction of excess graphene layers with respect to the as-grown material using our delamination method.

\subsection*{Type A epigraphene.}
\begin{figure} 
	\centering
	\includegraphics{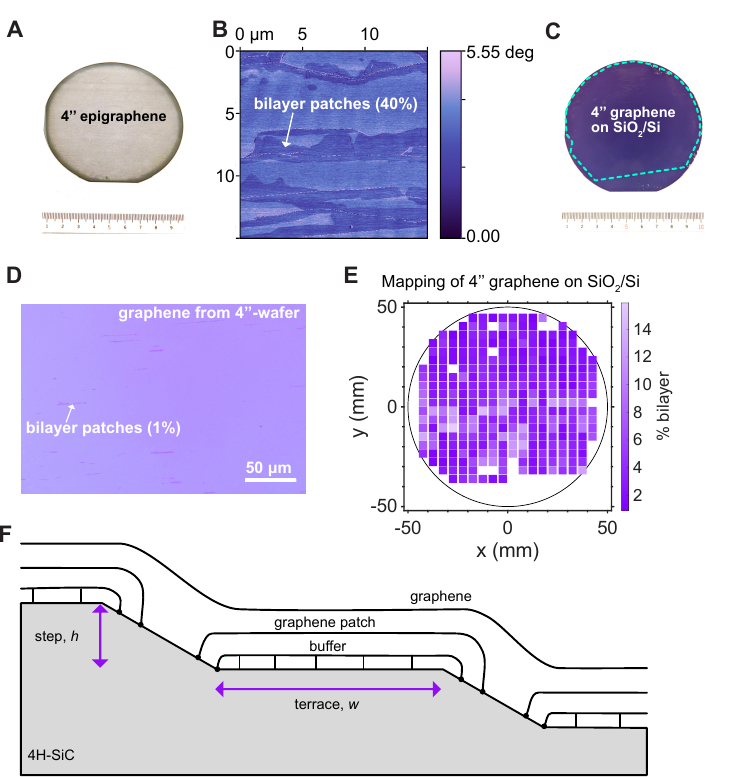} 
	\caption{\textbf{Delamination and transfer of wafer-scale epigraphene.}  (\textbf{A}) As-grown epigraphene on a recycled 4'' SiC wafer. (\textbf{B}) A representative 15$\times$15 $\mu$m$^2$ AFM phase map of epigraphene grown on a 4''-SiC wafer, showing the presence of bilayer domains (dark) that covers 40\% of the total area. (\textbf{C}) 4''-sized graphene transferred from SiC to a 4''-SiO$_2$/Si wafer. (\textbf{D}) Optical micrograph of graphene transferred from the 4''-SiC wafer to SiO$_2$/Si (C), showing the presence of bilayer domains (dark stripes) on monolayer graphene that cover 1\% of the area. (\textbf{E}) Mapping of percentage bilayer graphene over the SiO$_2$/Si wafer (see panel C) with the median value $4.3\%$. Each square corresponds to an area of $290 \times 190$ $\mu$m$^2$; the white regions correspond to areas with no graphene. (\textbf{F})  Schematic illustration of epigraphene on SiC within our growth model. The black circles represent anchor points of graphene patches or the buffer layer to the substrate. 
    }
	\label{fig:4} 
\end{figure}

Our experiments also provide an alternative perspective on epigraphene growth and the significance of SiC surface morphology for producing patch-free graphene. 
In addition to gold and silver as described earlier, we find that nickel -- reported to delaminate graphene multilayers from SiC and graphite \cite{Kim2013Layer-ResolvedLayers,Bae2017UnveilingGraphene,Moon2020Layer-engineeredGrapheneb} -- leads to highly selective monolayer delamination from a Type A substrate (see Fig. \ref{SI:AFM_Ni}). To interpret these results, we introduce a thermodynamic exfoliation model, where we use (i) free-energy differences to predict surface termination and structure \cite{Rohrer2010AbDeposition} and (ii) density functional theory (DFT) to model Si sublimation by tracking surface relaxations \cite{Rohrer2010AbDeposition,Giannozzi2017AdvancedESPRESSO}. We compute parameters for a total-system-energy descriptor using van der Waals (vdW)-inclusive DFT \cite{Giannozzi2017AdvancedESPRESSO,Dion2004VanGeometries,Hyldgaard2020ScreeningFoundation} (see Methods, Supplementary Text and Figs. \ref{SI_fig_DFT_2_stick_and_ball}-\ref{SI_fig_DFT_potential}). The Type A epigraphene in our model is illustrated in Figure \ref{fig:4}F: SiC terraces are covered with a buffer layer, patches are anchored to the substrate by chemical bonds to Si atoms at the step edges, and the top graphene layer extends over the whole substrate. The vdW attraction between graphene and the patch is slightly stronger than that between the patch and buffer, but weaker than the Si--C anchoring per carbon atom involved. Hence, our model predicts that wide patches are likely to co-delaminate with the top graphene layer, because as its width increases, the net impact by vdW forces outweighs the Si--C anchoring (see Fig. \ref{SI_fig_DFT_potential}). This gives an interpretation why a Type A surface (with its narrow patches) enables selective monolayer delamination.

In summary, we have shown that the ability to produce single-crystalline graphene, free from graphene multilayers and grain boundaries, stems from our capability to achieve a specific (Type A) surface reconstruction of 4H-SiC. We transferred the graphene monolayer from the SiC substrate without compromising its crystalline quality, resulting in high carrier mobilities and preserving all graphene-specific quantum transport features. Furthermore, by extending our process to recycled wafers, we demonstrate that the high cost associated with the SiC substrate \cite{Arvidsson2017ProspectiveMaturity} can be circumvented. Our results thereby establish SiC as an attractive generic platform -- alternative to CVD on metal catalysts -- for the mass production of single-crystalline graphene needed in high-performance applications. 


\begin{figure} 
	\centering
	\includegraphics{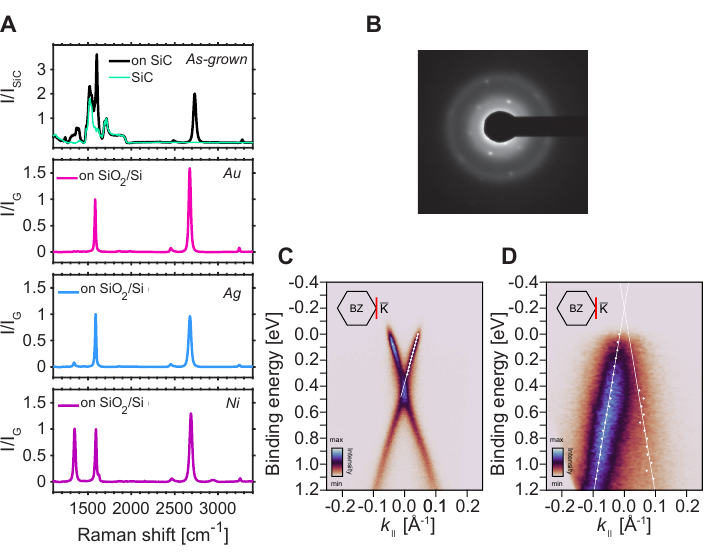} 

\caption{\textbf{Surface characterization of graphene delaminated from SiC.} (\textbf{A}) Raman spectra (532 nm laser) for as-grown epigraphene on SiC (black line) along with the 4H-SiC substrate underneath the graphene (green line), and graphene transferred from SiC to SiO$_2$/Si using different metals (gold, silver, and nickel). (\textbf{B}) SAED pattern of graphene transferred from SiC onto a holey carbon TEM grid. The hexagonal spot array represents graphene with lattice parameter $a=2.54$ $\pm 0.02$ Å. The diffuse halo is characteristic of amorphous carbon of the TEM grid. (\textbf{C}) ARPES band structure for as-grown graphene on SiC represented as an energy–momentum cut through the $\bar{K}$-point in the Brillouin zone as indicated in the inset. The dotted line is a linear fit to the band dispersion close to the Fermi level used to extract the Fermi velocity. (\textbf{D}) Same as (C) for graphene transferred to SiO$_2$/Si. As a consequence of the polarization of the light, only one branch of the band structure is visible \cite{Gierz2011IlluminatingGraphene}.
    }
	\label{fig:2} 
\end{figure}

\begin{figure} 
	\centering
	\includegraphics{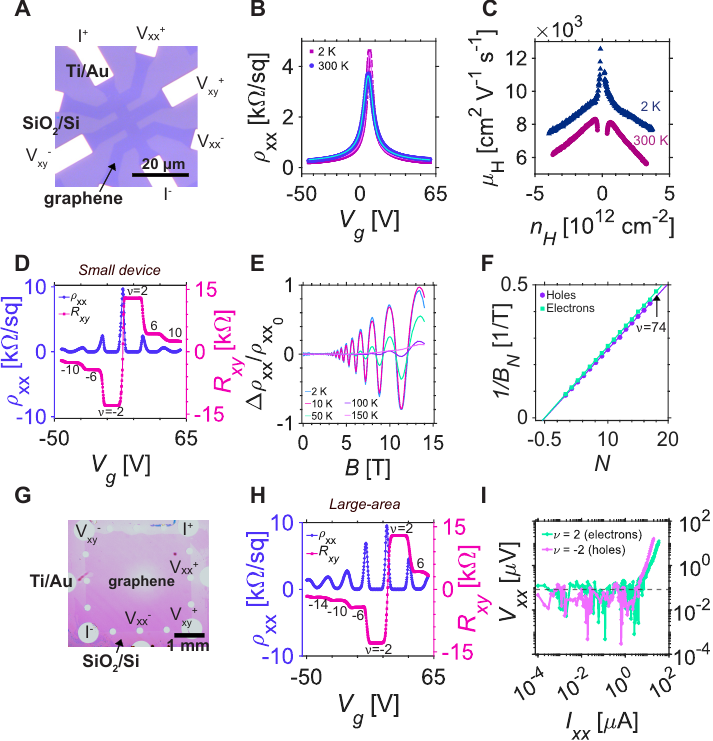} 
	\caption{\textbf{Electrical transport in graphene transferred from SiC onto SiO$_2$/Si.} (\textbf{A}) Hall bar with gold-transferred graphene ($W = 5\, \mu$m, $L = 15\,\mu$m) and passivated with PMMA. (\textbf{B}) Sheet resistivity vs. gate voltage at $T=300$ K (blue dots) and $T=2$ K (purple squares); solid lines are FET mobility fits. (\textbf{C}) Hall mobilities vs. carrier densities at $T=300$ K (blue dots) and $T=2$ K (magenta triangles). Solid lines are fits to Eq. \ref{mobility_vs_n_linearized}.
    (\textbf{D}) Half-integer QHE at $B=14$ T and $T=2$ K showing $\nu=2,6,10$. (\textbf{E}) Temperature dependence of SdH-oscillations for electrons ($V_g=60$ V; $n = 3.8\times10^{12}$ cm$^{-2}$). (\textbf{F}) Landau index plot from SdH $\rho_{xx}$-minima at 2 K for holes ($V_g = -44$ V) and electrons ($V_g = 60$ V). (\textbf{G}) Large-area device  with silver-transferred graphene on SiO$_2$/Si. (\textbf{H}) Half-integer QHE at $B=14$ T, $T=2$ K in the large-area device. (\textbf{I}) Zero-resistance state ($\rho_{xx}=0$) in I-V-curves measured at the $\nu  = \pm 2$-plateaus ($B=14$ T) for electrons ($V_g = 35$ V) and holes ($V_g = 15$ V) for $I_{xx}>0$. The dashed line indicates the noise floor ($V_{rms} = 83$ nV).
    }
	\label{fig:3} 
\end{figure}

\clearpage 

%
\bibliography{science_template} 

\bibliographystyle{sciencemag}

%
%
%
%
%
%


\section*{Acknowledgments}

We gratefully acknowledge Prof. Rositsa Yakimova and Prof. August Yurgens for  insightful discussions.

\paragraph*{Funding:}

This work was jointly supported by Chalmers Area of Advance Nano, Chalmers Area of Advance Materials, 2D-Tech VINNOVA competence Center (Ref. 2024-03852), and the Swedish Research Council VR under Contract No. 2021-05252 (J.H., S.L.A, N.S., J.J.P.), VR Contract No. 2022-03277 (P.H.), VR Contract No. 2020-04997 (E.S.), Knut and Alice Wallenberg Foundation No. 2018-0104 (O.T, M.H.B.), US National Science Foundation (NSF) award No. 2412500 (R.M-D., M.S.), VINNOVA 2023-03438 (H.H.), and VR Grant No. 2024-04485, Olle Engkvist Grant No. 233-0339, Nordita and KAW No. 2024.0129 (J.H.). Computer resources were supported by Chalmers C3SE and the National Academic Infrastructure for Supercomputing in Sweden under contracts NAISS2024/3-16 and NAISS2024-6-432 (E.S, P.H). This work was performed in part at Myfab Chalmers and Chalmers Materials Analysis Laboratory (CMAL).

\paragraph*{Author contributions:}

 J.H. (Johanna Huhtasaari), J.J.P., H.H., N.S., and S.L-A. performed experiments. J.H. (Johanna Huhtasaari), J.J.P., A.S., and N.S. contributed with growth of epigraphene material. M.H.B. and O.T. contributed with ARPES experiments. P.H. and E.S. contributed with modeling. M.S. and R.M-D performed electron diffraction (TEM) measurements. J.H. (Johannes Hofmann), T.B., N.S., and S.L-A. supervised the work. J.H. (Johanna Huhtasaari) and S.L-A. co-wrote  the manuscript with contributions from all authors. 

\paragraph*{Competing interests:}
The authors declare no competing interests.

\paragraph*{Data and materials availability:}
\subsection*{Supplementary materials}
Materials and Methods\\
Supplementary Text\\
Figs. S1 to S15\\
Tables S1 to S2\\


\newpage


\renewcommand{\thefigure}{S\arabic{figure}}
\renewcommand{\thetable}{S\arabic{table}}
\renewcommand{\theequation}{S\arabic{equation}}
\renewcommand{\thepage}{S\arabic{page}}
\setcounter{figure}{0}
\setcounter{table}{0}
\setcounter{equation}{0}
\setcounter{page}{1} 


\begin{center}
\section*{Supplementary Materials for\\ \scititle}


 \author{Johanna~Huhtasaari},
 \author{Joyal~Jain Palakulam},
 \author{Awse~Salha}, \\
\author{Per~Hyldgaard}, 
\author{Elsebeth~Schröder},       
\author{Magnus~Hårdensson~Berntsen},\\
\author{Oscar~Tjernberg},
\author{Manasi~Shah},
\author{Rodrigo~Martinez-Duarte},\\
\author{Hans~He}, 
\author{Johannes~Hofmann}, 
\author{Thilo~Bauch},
\author{Naveen~Shetty},\\
\author{Samuel~Lara-Avila}$^\ast$ \\
\small$^\ast$Corresponding author. Email: samuel.lara@chalmers.se\\
\end{center}

\subsubsection*{This PDF file includes:}
Materials and Methods\\
Supplementary Text\\
Figures S1 to S15\\
Tables S1 to S2\\


\newpage


\subsection*{Materials and Methods}\label{methods}




\subsubsection*{Growth of graphene}\label{method:graphene_growth}

Epitaxial graphene on silicon carbide (epigraphene) was grown in tight-confinement conditions by thermal decomposition of the silicon-terminated face (0001) of single-crystalline 4H-SiC. The samples were grown in an atmosphere of argon at 800 mbar, in the range of 1750$^\circ$C-1950$^\circ$C, for 5 minutes \cite{Virojanadara2008Homogeneous6H-SiC0001}. The monolayer coverage of the graphene samples was determined from transmission mode micrographs with gamma correction \cite{Yager2013ExpressTransport}. 
Graphene was grown on $7 \times 7$ mm$^2$-chips or 4''-sized wafers, 4'' being the larges size possible to fit inside our growth reactor. Graphene was grown three times on the same wafers by removing graphene with oxygen plasma after each growth; the best result is obtained by re-polishing wafers.

\subsubsection*{Transfer of graphene}\label{method:graphene_transfer}

Metal films (gold, silver or nickel) were deposited on the surface of as-grown epigraphene using an electron beam evaporator (Lesker PVD 225) at a base pressure of 5$\times10^{-7}$ Torr, with deposition rate $R=1$ Ås$^{-1}$. Metal thicknesses used for the experiments include 20, 50, 100 and 200 nm. Graphene delamination occurs in all cases, but thicker films provide better mechanical stability for successful transfer. After metal deposition, the samples were spin coated with PMMA (polymethylmethacrylate) resist (350 nm), and post-baked at $T=170^\circ$C for $t=5$ minutes. The polymer/metal/graphene layer was peeled off from the SiC substrate by hand using thermal release tape (Nitto Denko Corp.) with release temperature $T_{\textup{release}}=$ 120$^\circ$C or 170$^\circ$C.

\subsubsection*{Device fabrication}
Electrical devices were fabricated on graphene transferred onto SiO$_2$/Si (285 nm oxide thickness) using standard electron beam lithography, where electrical contacts were fabricated using physical vapor deposition of 5 nm Ti followed by 95 nm Au. For chip-sized devices, contacts were evaporated through a shadow mask. After fabrication of the devices, they were spin-coated with a layer of $>100$ nm PMMA as encapsulating layer before measurements. 

\subsubsection*{Atomic force microscopy of graphene}

Atomic force microscopy was performed in standard tapping mode or Peak Force tapping mode using a Bruker Dimension Icon AFM. The tip RTESP-300 was used (n-doped silicon) with resonance frequency $f_0 = 300$ Hz and spring constant $k=40$ N/m.

\subsubsection*{Optical mapping of graphene wafer}
A wafer of graphene on SiO$_2$/Si (Fig. \ref{fig:4}A) was optically mapped using a Nikon L200ND automatic microscope with a 50X objective, and the bilayer content on the graphene was determined using image post-processing (tresholding using e.g. the software ImageJ).

\subsubsection*{Electrical measurements}

Electrical measurements were carried out using a Physical Property Measurement System (PPMS) from Quantum Design, at temperatures from $T=300$ K to 2 K, and magnetic fields up to $B=14$ T. Samples were biased with a DC current of 10 nA or 100 nA for small Hall bars, and 100 nA-30 $\mu$A for large (5 mm $\times$ 5 mm) graphene devices using a Keithley 6211 current source, and sample voltages were measured using an Agilent 34420 A nanovolt-meter. The electrostatic bottom gate was DC-biased in the range $-50\,\mathrm{V}\leq I \leq 60\, \mathrm{V}$ using a Keithley 2400 source meter. All electrical measurements were carried out in four-probe configuration. In gated measurements, leakage currents were kept below 0.6 nA. The noise floor in the zero-resistance state in Figure \ref{fig:3}I was calculated as the RMS of the voltage in the flat ($I < 3 \mu$A) region. 

\subsubsection*{Raman spectroscopy} 

Raman spectroscopy was carried out using a WITec alpha300 R Raman microscope with a 532 nm laser and 600 g/mm grating. Raman spectra of graphene on SiO$_2$/Si were normalized by the intensity of the G-peak, at $\sim 1600$ cm$^{-1}$, and graphene spectra of pristine epigraphene on SiC were normalized to the peak at 1714 cm$^{-1}$ which represents the SiC substrate. The spectrum of the SiC substrate was obtained focusing 20 $\mu$m down into the substrate on the same spot as the graphene spectrum. 

\begin{figure} 
	\centering
	\includegraphics{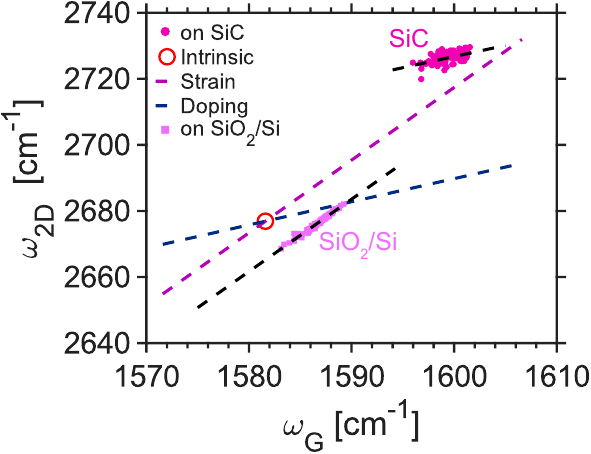} 
	\caption{\textbf{Detailed strain/doping Raman analysis for epigraphene (on SiC) and transferred to SiO$_2$/Si.} The 2D-peak positions ($\omega_{2D}$) vs. G-peak positions ($\omega_{G}$) for graphene on SiC (magenta dots), and graphene transferred to SiO$_2$/Si (pink squares). Black dashed lines are linear fits to each data set. Purple and blue dashed lines represent expected behavior for strain- and doping dominated graphene \cite{Lee2012OpticalGraphene}; red ring represents suspended graphene at 514 nm laser wavelength \cite{Lee2012OpticalGraphene}. The Raman correlation of epigraphene follows the doping trajectory ($\Delta \omega_{2D}/\Delta\omega_G \approx 0.7$), indicating charge transfer–dominated behavior. The data points for graphene on SiO$_2$/Si lie in the quadrant Q4 as defined in Lee et al. \cite{Lee2012OpticalGraphene}, indicating tensile strain and p-doping. 
    }
	\label{supplementary:Raman_strain_doping_analysis_PLOT} 
\end{figure}

For the detailed Raman analysis in Figure \ref{supplementary:Raman_strain_doping_analysis_PLOT}, the 2D- and G-peak positions were extracted from large-area scans of 100 spectra in areas  of $27.1 \times 26.8$ $\mu$m$^2$ for graphene on SiO$_2$/Si and $41.6 \times 45.2$ $\mu$m$^2$ for as-grown epigraphene using single-Lorentzian fits to 2D- and G-peaks. The shifts of the Raman 2D- and G-peak positions relative to the intrinsic graphene point ($\omega_{G,0}, \omega_{2D,0}$) were decomposed into contributions from mechanical strain and charge transfer (doping) following Lee et al \cite{Lee2012OpticalGraphene}.

\subsubsection*{Selected area electron diffraction}\label{method:TEM}

Selected Area Electron Diffraction (SAED) measurements were performed using a transmission electron microscope (TEM) operated at an acceleration voltage of 300 keV. The electron wavelength ($\lambda$) corresponding to this voltage ($V$) was calculated using the relativistic equation:

\begin{equation}
    \lambda = \frac{h}{\sqrt{2meV(1+\frac{eV}{2m_ec^2})}}
\end{equation}
where $h$ is Planck's constant, $m_e$ the free electron mass, and $c$ the speed of light. For 300 kV, the calculated wavelength was 0.0196 Å. 

The interplanar spacing $d_{hk}$ was obtained from the SAED pattern using the relation for a 2D hexagonal lattice:

\begin{equation}
    d_{hk} = \frac{\lambda L}{R},
\end{equation}
where $L$ is the camera length ($0.25$ m) and $R$ is the measured distance from the central beam to the diffraction spot. $R$ was determined from the SAED image by measuring the distance to the (100) diffraction spot multiple times to obtain an average value, and the measured $R_{100}$ gave $d_{100} = 2.20$ Å. 

The lattice constant ($a$) was calculated from the d-spacing via the relation for a 2D hexagonal lattice:

\begin{equation}
    d_{hk} = \frac{a \sqrt{3}}{2\sqrt{h^2+k^2+hk}} \Rightarrow \\ d_{100} = \frac{a \sqrt{3}}{2} 
\end{equation}
The error was estimated as the standard deviation of multiple $R_{100}$ measurements.

\subsubsection*{Angle-Resolved Photoemission Spectroscopy}

Angle-Resolved Photoemission spectroscopy (ARPES) measurements were carried out in ultra-high vacuum  at room temperature. The sample was illuminated by 18.1 eV photons provided by a femtosecond high-harmonic generation (HHG) light source \cite{Guo2022ASpectroscopy}.

\subsubsection*{Field-Effect mobility fits}\label{supplementary:FET} 

\begin{figure} 
	\centering
	\includegraphics{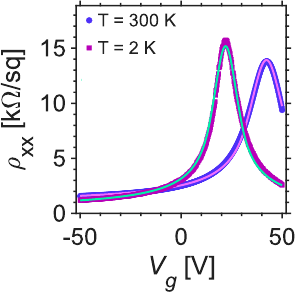} 
	\caption{{\textbf{Gate voltage-response of the 5 $\times$ 5 mm graphene device on SiO$_2$/Si.} Sheet resistivities were estimated from van der Pauw-measurements that give $L/W\approx0.4$ squares. Solid lines are fits to extract FET mobilities (\ref{eq:S4}): $\mu= 1,100$ cm$^2$V$^{-1}$s$^{-1}$ at $T = 300$ K, and  $\mu=1,300$ cm$^2$V$^{-1}$s$^{-1}$ at $T=2$ K.} 
    }
	\label{supplementary_fig:largemobility_RvsVg} 
\end{figure}

We extracted the field-effect mobility in Figs. \ref{fig:3}B and \ref{supplementary_fig:largemobility_RvsVg} from a fit of the longitudinal resistivity $\rho_{xx}$ ($B=0$ T) to a model that accounts for spatial fluctuations in the carrier density (charge puddles) near the Dirac point \cite{Gosling2021UniversalImpurities}. Three fitting parameters were used: the mobility $\mu_{FET}$, the carrier density broadening $\delta n$, and the residual resistivity $\rho_s$. 
The effective carrier density was calculated from the applied gate voltage as 

\begin{equation}
    n_{g} = \frac{C_0}{e}\,(V_g - V_D),
\end{equation}
where $C_0$ was taken as the capacitance for a 285 nm SiO$_2$ gate oxide, $C_0 = 121$ $\mu$Fm$^{-2}$.

To account for charge density broadening around charge neutrality, the carrier density was defined as a quadratic function near the Dirac point and otherwise as a linear function \cite{Gosling2021UniversalImpurities}:

\begin{equation}
    n(V_g) =
\begin{cases}
\frac{\delta n}{4} + \frac{n_{\mathrm{g}}^2}{\delta n}, & |n_{\mathrm{g}}| \leq \delta n/2 \\[10pt]
|n_{\mathrm{g}}|, & |n_{\mathrm{g}}| > \delta n/2.
\end{cases}
\end{equation}
The carrier density was used in the resistivity expression

\begin{equation}\label{eq:S4}
\rho(V_g)  = \dfrac{1}{e \mu_{FET} \, n(V_g)} + \rho_s
\end{equation} 
and the fitting parameters $\mu$, $\delta n$, and  $\rho_s$ were determined by fitting the measured $\rho_{xx}$ vs. $V_g$ to Eq. \ref{eq:S4}.

The density of charged impurities was estimated from the fitted mobility value ($\mu = \mu_{FET}$ as

\begin{equation}\label{n_imp}
    n_{\mathrm{imp}} = (50\, n_0 / \mu)\mu_0 
\end{equation} 
with $n_0 = 10^{14}$ cm$^{-2}$ and $\mu_0 = 10^{4}$ cm$^{2}$V$^{-1}$s$^{-1}$  \cite{Adam2007ATransport}.

\subsubsection*{Fits of Hall mobility vs. density}\label{supplementary:Hall_fits} 

The Hall mobility vs. carrier density in Figure \ref{fig:3}C was obtained via measurements of the longitudinal ($\rho_{xx}$) and transverse ($R_{xy}$) resistance components vs. back-gate voltages at a constant perpendicular magnetic fields ($B=-1$ T, $B=0$ T and $B=1$ T). Data corresponding to the charge puddle regime \cite{Zhang2009OriginGraphene} close to charge neutrality ($n_H = 0$), where $R_{xy}(B)$ is non-linear, were omitted.

At each gate voltage, the Hall coefficient was calculated as 

\begin{equation}
   R_H =   \frac{\mathrm{d}R_{xy} }{\mathrm{d} B}.
\end{equation}
The carrier densities $n_H$ were calculated as

\begin{equation}\label{n_H}
    n_H =  \frac{1}{R_H e}
\end{equation}
with $e$ the elementary charge ($e>0$) such that $n_H <0$ for holes and $n_H > 0$ for electrons.

The Hall mobilities were calculated as

\begin{equation}\label{hall_mobility}
   \mu_H = \frac{\vert R_H \vert }{\rho_{xx }(B=0\,  \textup{T}) }.  
\end{equation}

The measured Hall mobility ($\mu_H$) vs.carrier density was fitted to the carrier density ($n_H$) using a model that combines charged impurity and short-range scattering. Via Matthiessen’s rule, the total mean free path is given by:
\begin{equation}
\frac{1}{l_{\mathrm{mfp}}} = \frac{1}{l_{\mathrm{mfp}}^{\mathrm{charge}}} + \frac{1}{l_{\mathrm{mfp}}^{\mathrm{short}}},
\end{equation}
with \cite{Nomura2007QuantumFermions}
\begin{equation}
l_{\mathrm{mfp}}^{\mathrm{charge}} \approx A\sqrt{n}, \qquad l_{\mathrm{mfp}}^{\mathrm{short}} \approx \frac{B}{\sqrt{n}}. 
\end{equation}

This yields
\begin{equation}
l_{\mathrm{mfp}} = \frac{A B \sqrt{n}}{B + A n}.
\end{equation}
Using the semiclassical relation for Dirac fermions,
\begin{equation}
\mu = \frac{e\,l_{\mathrm{mfp}}}{\hbar \sqrt{\pi n}},
\end{equation}
the mobility becomes
\begin{equation}
\mu_H(n) = \frac{C}{1 + (A/B)n},
\end{equation}
where \( C = (50\, n_0 / n_{\mathrm{imp}})\mu_0 \) ($C$ is called $\mu$ in Eq. \ref{n_imp}) \cite{Adam2007ATransport}. This was linearized as:

\begin{equation}\label{mobility_vs_n_linearized}
\mu \approx C \left(1 - \frac{A}{B}n\right)
\end{equation}
for small $(A/B)n$. Using $C$ as a fitting parameter, a fit of the Hall mobility $\mu = \mu_H$ (Eq. \ref{hall_mobility}) vs. carrier density $n=n_H$ (Eq. \ref{n_H}) allowed to estimate $n_{\mathrm{imp}}$ for the Hall measurements.

\subsubsection*{Temperature dependence of Shubnikov-de Haas-oscillations}

The temperature dependence of the Shubnikov–de Haas (SdH) oscillations in Figure \ref{fig:3}E ($\rho_{xx}$ vs.$B$) was analyzed within the Lifshitz–Kosevich (LK) formalism \cite{Ando1982ElectronicSystems,Tan2011Shubnikov-deBias}. The $\rho_{xx}$ vs.$B$-data at each constant temperature $T$ were normalized as

\begin{equation}
    \Delta \rho_{xx}/\rho_{xx,0} = \frac{\rho_{xx} - \rho_{xx}(B=0\,\textup{T})}{ \rho_{xx}(B=0\,\textup{T})} 
\end{equation}
and the oscillations were isolated from the $T = 2, 4, 10, 20, 50$ K-data by subtracting a nearly linear (spline) background fitted to the oscillation's center-line; for the $100$ K-data a second-order polynomial was subtracted.
The SdH oscillation amplitudes $A_i$ were estimated as the $\Delta \rho_{xx}/\rho_{xx,0}$-values at oscillation minima. The uncertainties in $A_i$ -- $\sigma_{A_i}$ -- were estimated from the variance 
across different choices of background subtractions, combined in quadrature with an 8$\%$ error consisting of a conservative 5$\%$ experimental error and a 3$\%$ error for differences in carrier densities at temperatures (because the effective mass depends on the carrier density for Dirac fermions). 

\begin{figure} 
	\centering
	\includegraphics{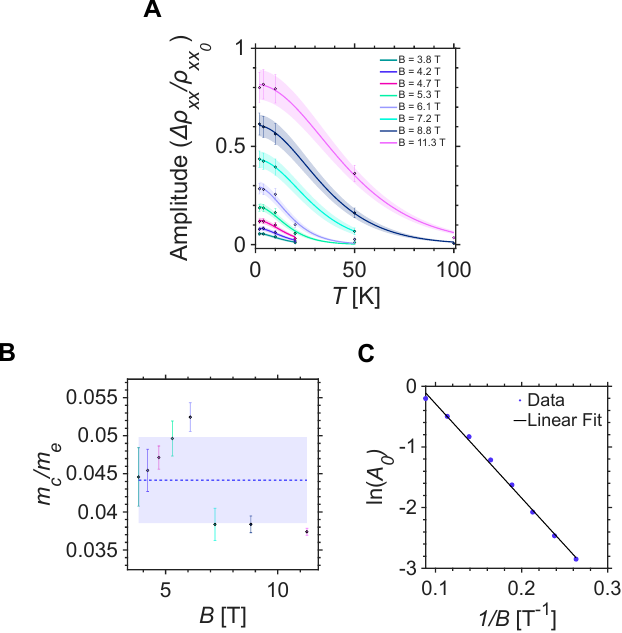} 
	\caption{\textbf{Fits of SdH-oscillation amplitude to the Lifshitz-Kosevich formula. } (\textbf{A}) Temperature dependence of the SdH amplitudes fitted to the Lifshitz–Kosevich model (Eq. \ref{LK_thermal}) to extract the cyclotron mass and SdH amplitude prefactors $A_0$. The shaded areas indicate the combined uncertainty from background subtraction and experimental errors (see Methods).
    (\textbf{B}) Cyclotron mass normalized to free-electron mass (obtained in (A)) vs. magnetic field strengths. The dashed line is the mean, $m_c/m_e=0.044 \pm 0.0056$ 
    and the shaded areas is the standard deviation. (\textbf{C}) Dingle plot showing $\ln (A_0(B_i))$ vs. $1/B_i$ at SdH extrema i. The slope of the linear fit provides the Dingle temperature ($T_D = 23.9 \pm 3.10$ K) and corresponding lifetime ($\tau = 50.8 \pm 6.59$ fs).
    } 
	\label{SI_sdH_temp_dependence_fig} 
\end{figure}

For each field $B_i$ corresponding to an oscillation minimum, the amplitudes $A_i(T)$ were fitted in Figure \ref{SI_sdH_temp_dependence_fig}A to:

\begin{equation}\label{LK_thermal}
A(T, B) = A_0(B) \frac{X}{\sinh X}
\end{equation}
with $X$: 
\begin{equation}\label{X}
X = \frac{2\pi^2 k_B T m_e (m_c/m_e)}{\hbar e B},
\end{equation}
where $k_B$ is the Boltzmann constant, $e$ the elementary charge, and $m_e$ the free-electron mass, and fitting parameters $A_{0,i}$ and $m_{c,i}/m_e$ (cyclotron mass normalized by free-electron mass). 

The uncertainties in $m_c/m_e$ (error bars in Fig. \ref{SI_sdH_temp_dependence_fig}B) were estimated from fits to the upper and lower amplitude curves defined by $A_i(T)\pm \sigma_{A_i}$:

\begin{equation}
    \Delta m_{c,i} = \frac{m_{c,i}^{upper} - m_{c,i}^{lower}}{2}.
\end{equation}

The mean cyclotron mass was calculated as the arithmetic mean of all fitted values and its uncertainty estimated as the standard deviation of all individual $m_{c,i}/m_e$ values.
The prefactor $A_0(B)$ describes the magnetic-field dependent damping due to a finite quantum lifetime $\tau$, and was written as \cite{Tan2011Shubnikov-deBias} 

\begin{equation}\label{A0}
A_0(B) = A_0^* \exp
\left(-\frac{\pi}{\omega_c \tau}\right),
\end{equation}

where the cyclotron frequency is
\begin{equation}\label{omegc}
\omega_c = \frac{eB}{m_c},
\end{equation}

and the lifetime is related to the Dingle temperature $T_D$ via
\begin{equation}\label{tau}
\tau = \frac{\hbar}{2\pi k_B T_D}.
\end{equation}

Combining these expressions yields
\begin{equation}\label{A0_complete}
A_0(B) = A_0^*
\exp\left[-\frac{2\pi^2 k_B T_D m_c}{e\hbar B}\right]
\end{equation}

with the logarithmic form

\begin{equation}\label{A0_log}
\ln A_0(B) = \ln A_0^* - \frac{2\pi^2 k_B T_D m_c}{e\hbar} \frac{1}{B}.
\end{equation}

A linear fit of $\ln A_0$ vs. $1/B$ (Fig. \ref{SI_sdH_temp_dependence_fig}C) provides the Dingle temperature $T_D$ from the slope, and the corresponding quantum lifetime from Eq.\ref{tau}.

The uncertainty in the Dingle temperature $T_D$ was estimated using standard Gaussian error propagation, combining the standard error of the fitted slope of $\ln A_0$ versus $1/B$ with the uncertainty in the effective mass. The uncertainty in the lifetime was obtained by propagating $\sigma_{T_D}$ through Eq. \ref{tau}.

\subsubsection*{Landau index plots}
The Landau level index $N$ was assigned to the Shubnikov--de Haas (SdH) oscillation minima in $\rho_{xx}$. Plotting $1/B_N$ versus $N$ gives a linear relation
\begin{equation}\label{ns}
\frac{1}{B_N} = \frac{4e}{h n_s}\left(N + \tfrac{1}{2}\right),
\end{equation}
from which the slope yields the carrier density $n_s$.

\subsubsection*{Thermodynamic modeling of exfoliation}

We use thermodynamic modeling of externally driven transition processes \cite{Rohrer2010AbDeposition} to extract predictions for the probability of selective exfoliation as a function of the characteristic terrace width. The model reflects our picture of selective Si sublimation and graphene-plus-patch formation that is summarized in the main text and developed and further described in Figure \ref{fig:4}F. We treat exfoliation of any given section of a graphene as an example of a surface-termination problem where the outcome 
is dependent on the environment (namely outcomes for other graphene segments) but set by a competition between metastable configurations (see inserts of Fig. \ref{SI_fig_DFT_energy_diagram2} and \cite{Rohrer2010AbDeposition}). In practice, our predictions emerge after asserting a total-system-energy form (Fig. \ref{SI_fig_DFT_potential}) from DFT inputs (see Supplementary Text, DFT-method subsection, and Figs. \ref{SI_fig_DFT_2_stick_and_ball}-\ref{SI_fig_DFT_energy_diagram2}) and setting the ratio of outcome probabilities (for the investigated graphene segment) to a Gibbs-free-energy weighting \cite{Rohrer2010AbDeposition}.

\subsubsection*{Density Functional Theory}

Density functional theory predictions of motifs in the selective Si sublimation, of the strength of Si-C anchor bonds, and (for illustration) of the delamination were done in Quantum Espresso \cite{Giannozzi2017AdvancedESPRESSO}, using ultrasoft pseudopotentials with the exchange-correlation approximation set by the strictly parameter-free consistent-exchange vdW-DF-cx \cite{Dion2004VanGeometries,Hyldgaard2020ScreeningFoundation}. We used a slab geometry with a 4$\times$4$\times$3 repetition of the 4H-SiC unit cell and tracked both surface relaxations and structure changes upon Si atom sublimation.



\subsection*{Supplementary Text}



\subsubsection*{ARPES}

\begin{figure} 
	\centering
	\includegraphics{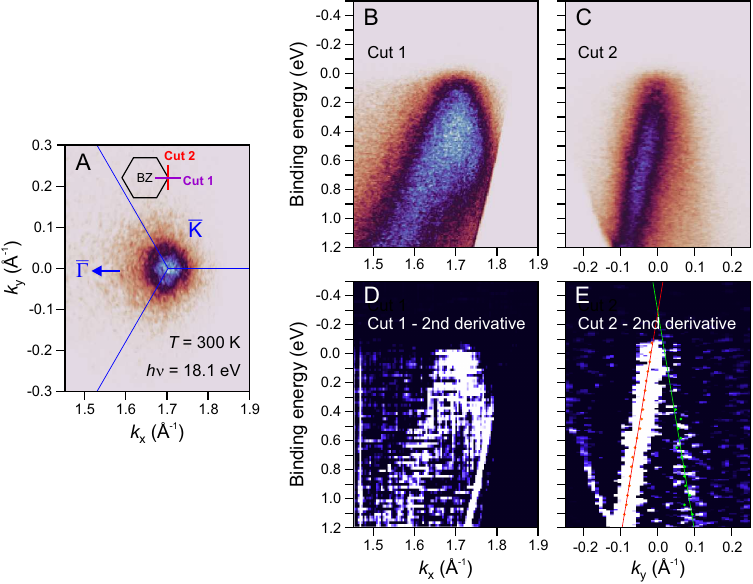} 
	\caption{\textbf{ARPES measurements on epigraphene transferred onto a SiO$_2$/Si substrate.} (\textbf{A}) Fermi surface of graphene at room temperature near the $\bar{K}$ point. (\textbf{B}) and (\textbf{C}) Energy-momentum cut through the directions indicated in the inset of panel A.  (\textbf{D}) and (\textbf{E}) Second-derivative plots of the data in panels A and B, respectively. The red (green) dots mark the fitted peak positions for the left (right) branch of the Dirac cone based on the intensity plot in panel C. Solid lines represent linear fits to the extracted peak positions.  
    }
	\label{SI:ARPES_fig} 
\end{figure}

Figure \ref{SI:ARPES_fig} presents room-temperature ARPES data from delaminated graphene transferred onto a SiO$_2$/Si substrate. Panel A shows the Fermi surface near the $\bar{K}$ point of the Brillouin zone (BZ). Panels B and C display the energy-momentum dispersion along the momentum directions defined by Cut 1 and Cut 2 in the BZ inset of Panel A. In graphene, the measured photoemission intensity is highly sensitive to the polarization of the incident light \cite{Gierz2011IlluminatingGraphene}, leading to the well-known suppression of one branch of the Dirac cone under certain experimental geometries. This explains why the right branch of the cone appears missing in Panel C. 

To enhance band dispersion visibility, two-dimensional second derivatives of the spectra in Panels B and C were computed, and the results are shown in Panels D and E. In Panel E, the right branch of the Dirac cone becomes discernible despite its weak intensity. Momentum distribution curves (MDCs) extracted from Panel C at higher binding energies ($E_b$) further confirm this dispersion. Fitted MDC peak positions of the left and right branches are shown as red and green dots in Panel E, overlaid on the second-derivative plot. Linear fits ($ak + b$) to these data yield Fermi velocities $v_F = \vert a \vert e /\hbar \times 10^{-10}$ ms$^{-2}$ where $a$ is given in units of eVÅ$^{-1}$ and $e=1.6\times10^{-19}$ C, giving $v_F = 2.20\times 10^6$ ms$^{-1}$ and $v_F = 2.18\times 10^6$ ms$^{-1}$ for the left and right branches, respectively. 

Comparing Panel C with the corresponding spectrum for as-grown graphene on SiC (Fig. \ref{fig:2}C, main text) reveals a significant momentum broadening after transfer to SiO$_2$/Si. Lorentzian fits to MDCs at the Fermi energy ($E_b = 0$ eV) give FWHMs of 0.013 Å$^{-1}$ (as-grown) and 0.105 Å$^{-1}$ (transferred). This nearly tenfold increase is expected based on previous reports for graphene on quartz \cite{Hwang2012FermiModification} and is attributed to enhanced electron-electron interactions due to reduced dielectric screening. The broadening thus reflects increased electron self-energy rather than impurity or contamination scattering, providing additional evidence of the high crystalline quality of the transferred graphene.

\subsubsection*{Density functional theory}
\begin{figure} 
	\centering
	\includegraphics{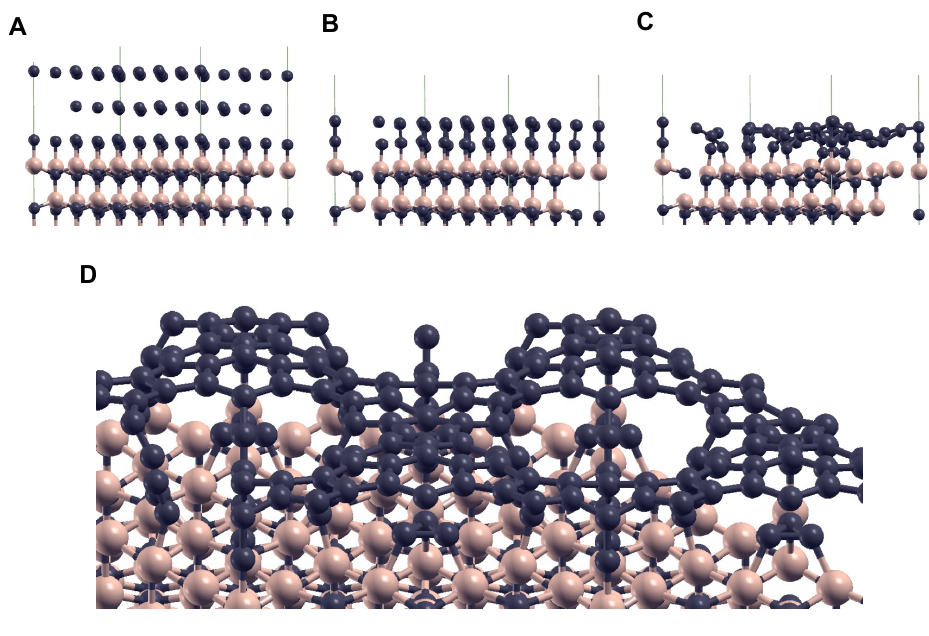} 
	\caption{{\textbf{Representative atomistic motifs from DFT simulations of a $4\times4$ in-plane extension of a 4H-SiC(0001).}
  } (\textbf{A}) Starting from a model with three Si layers removed, relaxation with a parameter-free functional \cite{Dion2004VanGeometries,Hyldgaard2020ScreeningFoundation} shows that 50 sp$^2$-hybridized C atoms cannot form both a buffer and a full graphene overlayer. Instead, Si atoms act as anchors for a partially arched patch over a small buffer island. (\textbf{B})-(\textbf{C}) Early stages show C covering exposed Si, but further relaxation leads to C clustering and selective Si–C anchoring.  (\textbf{D}) The final relaxed structure exhibits separation between a patch-like layer and buffer-like clusters, illustrating elements of Fig. \ref{fig:4}F and model-based predictions for patch-free graphene production (Figs. \ref{SI_fig_DFT_energy_diagram1}-\ref{SI_fig_DFT_potential}).}
	\label{SI_fig_DFT_2_stick_and_ball} 
\end{figure}

In the DFT studies, to get representative motifs and thus set model parameters, we employ a slab system that has a (0001) surface and is defined by a 4$\times$4$\times$3 repetition of the 4H-SiC cell but also permitted full atomic relaxations. From this slab, we selectively remove the topmost 48 Si atoms (see the top-left panel of Figure \ref{SI_fig_DFT_2_stick_and_ball}). The remaining panels of Figure \ref{SI_fig_DFT_2_stick_and_ball} show representative atomistic motifs that emerge as we next implement relaxations driven by the forces we compute in our vdW-inclusive DFT \cite{Dion2004VanGeometries,Hyldgaard2020ScreeningFoundation}.

\begin{figure} 
	\centering
	\includegraphics[width=0.8\textwidth]{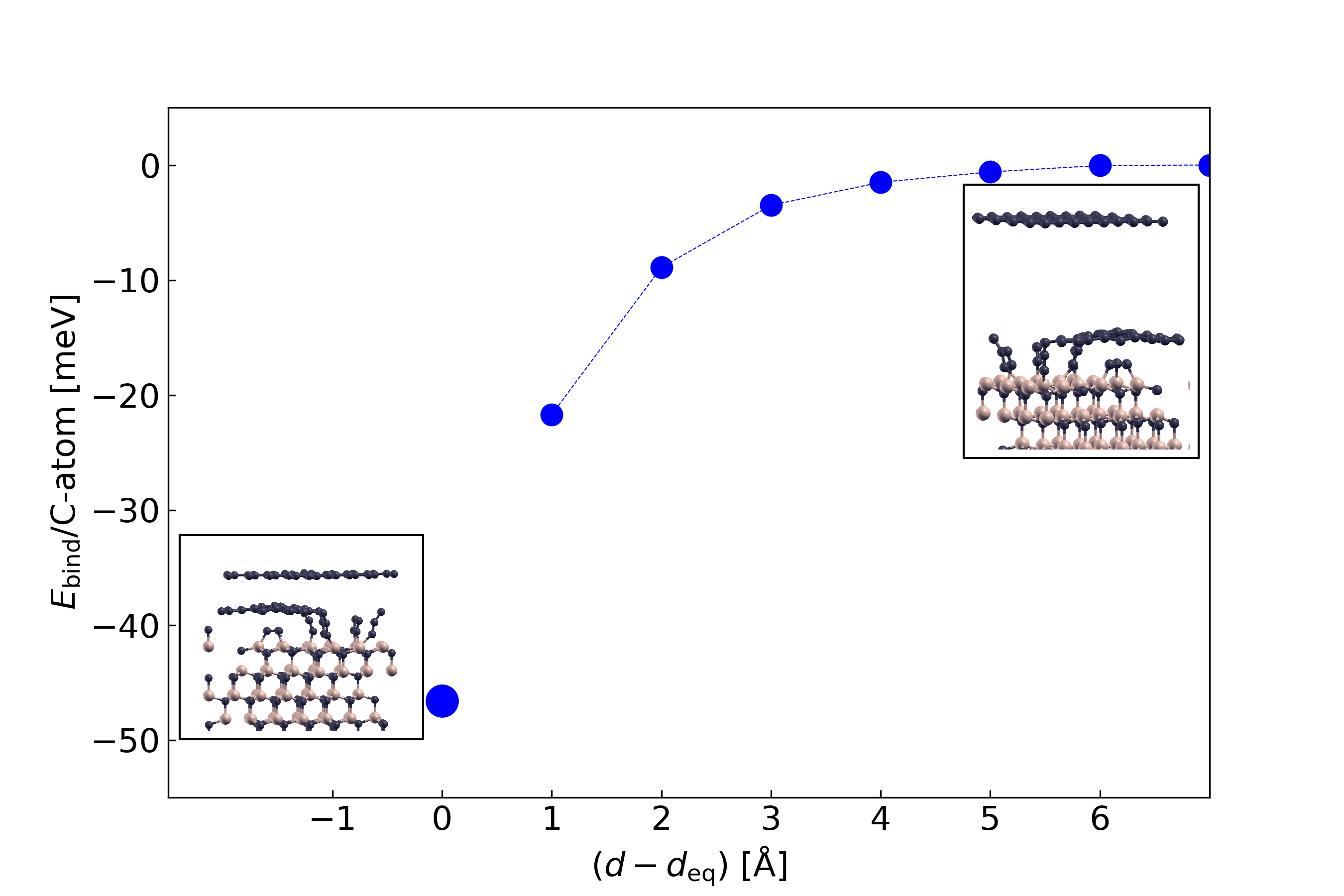} 
\caption{{\textbf{Direct DFT illustration that narrow terraces leads to selective exfoliation, i.e., that patches prefer to stay with the substrate and buffers for such SiC surfaces.}} This DFT illustration results after we add (and relax to optimal separation $d_{\rm eq}$ from the substrate) an extended graphene layer (see left-most insert) on top of the ground-state structure motif (Fig. \ref{SI_fig_DFT_2_stick_and_ball}) that we found emerging from DFT-computed forces. We compute the system energy (blue dots) as as a function of increasing graphene separation, $d-d_{\rm eq}$ (right-most insert) while implementing relaxations (again by DFT-computed forces) for all but one of the graphene atoms. Note that while this study nominally represents a system with just a 1.2 nm terrace width, it also characterizes the vdW-dominated graphene-patch adhesion for use in our thermodynamic modeling.}
	\label{SI_fig_DFT_energy_diagram1} 
\end{figure}

\begin{figure} 
	\centering
	\includegraphics[width=0.8\textwidth]{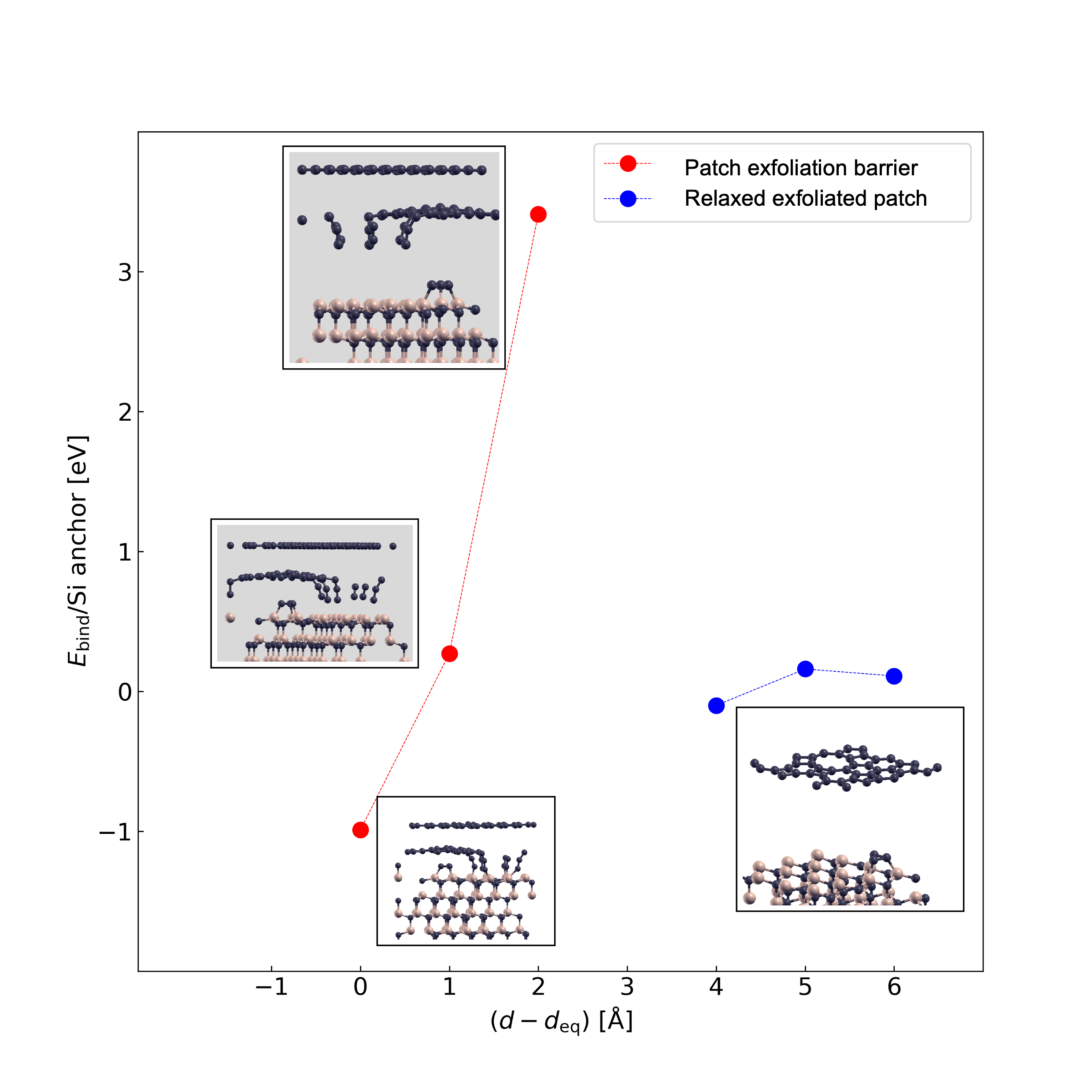} 
\caption{{\textbf{DFT characterization of exfoliation barriers and patch relaxation.}
  } Insets with gray backgrounds (red dots) show the nature and strength of the exfoliation barrier ($\approx 4.4$ eV per anchoring Si atom) in the presence of a graphene overlayer. Insets with white backgrounds illustrate the nature and energy of a patch displaced beyond abrupt Si–C bond breaking, relaxing to a DFT-optimized alignment with the top graphene layer. All calculations assume a given increase $d-d_{\rm eq}$ in the separation between a fixed graphene atom and the substrate, relative to the fully relaxed anchored state (bottom-most inset), but most atoms are free to relax under DFT forces. For abrupt Si-anchor breaking (red-dot cases), vertical positions of patch atoms are fixed at the given $d-d_{\rm eq}$. Beyond the transition ($d-d_{\rm eq}\approx 2.5$ {\AA}), full relaxation by DFT-computed forces are instead implemented, and what is initially a heavily strained patch (blue-dot cases) recovers an essentially flat geometry (shown in the right-most inset).}
	\label{SI_fig_DFT_energy_diagram2} 
\end{figure}

We note that, in the chosen unit cell, it takes 50 sp$^2$-hybridized carbon atoms to form a nearly relaxed graphene-type overlayer. The structure shown in Figure \ref{SI_fig_DFT_2_stick_and_ball}A does not have a sufficient number of unbound C atoms that our direct DFT modeling can ever produce a graphene-overlayer system like that of Figure \ref{fig:4}F. We turn this observation of a formal limitation of our motif characterization, Figure \ref{SI_fig_DFT_2_stick_and_ball}, into an advantage: The final, fully relaxed motif, permits us to both illustrate the exfoliation (Fig. \ref{SI_fig_DFT_energy_diagram1}) and complete DFT characterizations of the strength and nature of patch-to-graphene adhesion and of edge-located Si-C anchoring (Fig. \ref{SI_fig_DFT_energy_diagram1} and Fig, \ref{SI_fig_DFT_energy_diagram2}, respectively).

\begin{figure} 
	\centering
	\includegraphics[width=0.8\textwidth]{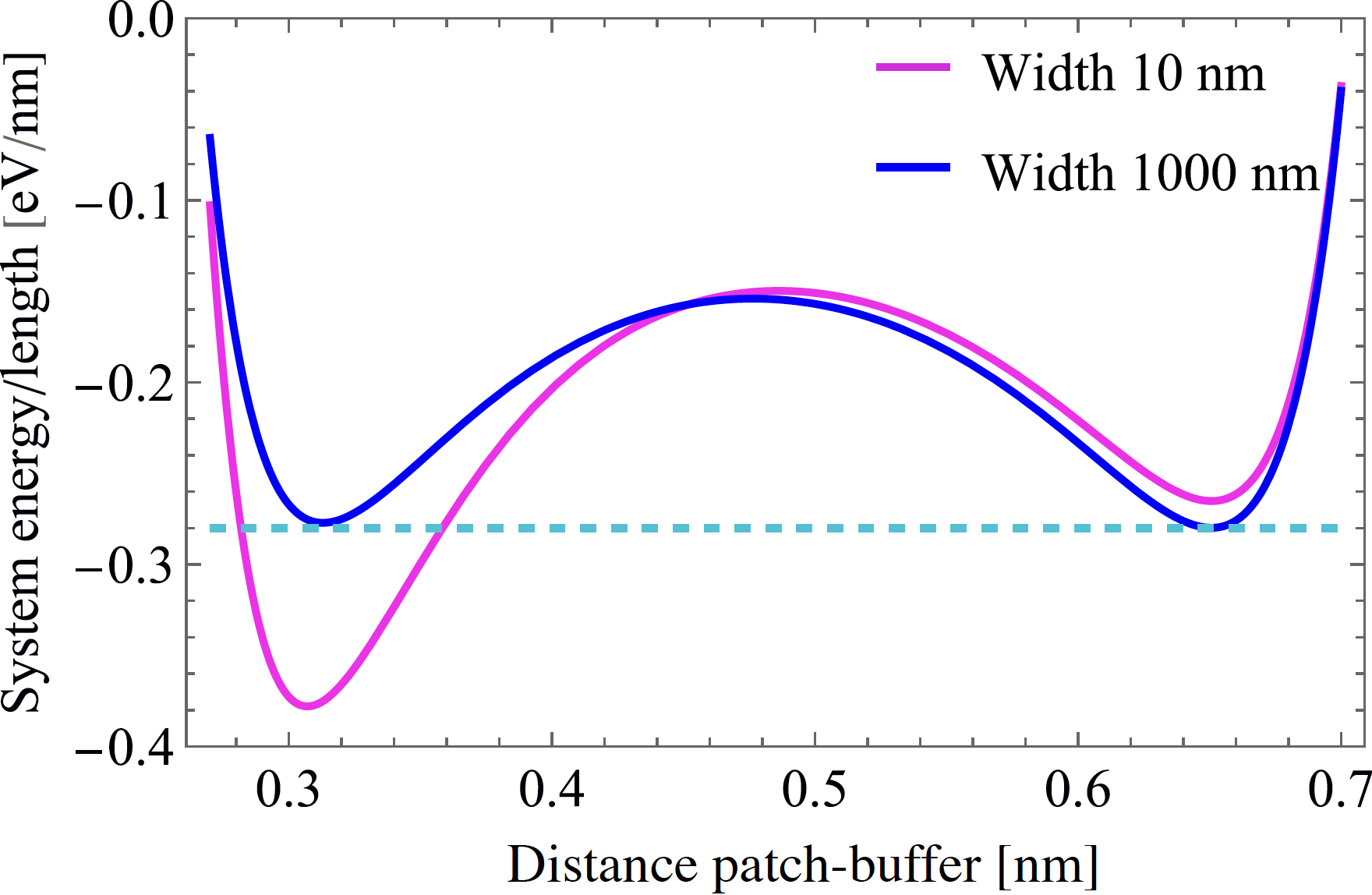} 
	\caption{{\textbf{Crossover with terrace width in our qualitative descriptor of likelihood for selective exfoliation.}} We assume a fixed, 1 nm separation between the graphene and the buffer and track the variation with the assumed patch-buffer separation in the total system energy-approximation used in our thermodynamic modeling. The left- (right)-most minima (each corresponding to the stated assumption of 
    the terrace width) therefore correspond to positioning the patch in vdW binding to the buffer (to the exfoliated graphene). The energy difference 
    between the minima is a measure of crossover in expected outcome \cite{Rohrer2010AbDeposition} and there is a clear
 crossover to a non-selective exfoliation with increasing terrace width.}
	\label{SI_fig_DFT_potential} 
\end{figure}

For the thermodynamic modeling we need a total-system-energy form \cite{Rohrer2010AbDeposition} that in our case depends on the separation 
$d$ between the graphene and the buffer as well as the patch-to-buffer separation (see Fig. \ref{SI_fig_DFT_potential}). The system-energy form
is naturally given as an average across the assumed width (and per length) of the patch. It comprises a patch-to-buffer-and-substrate binding energy `$E_1$' and the competing patch-to-graphene binding energy `$E_2$', with changes in the $E_1-E_2$ having exponential impact on the expected exfoliation outcomes \cite{Rohrer2010AbDeposition}. 
Figure \ref{SI_fig_DFT_energy_diagram1} provides a characterization of the $E_2$ terms that grow linearly with the assumed terrace width. Figure \ref{SI_fig_DFT_energy_diagram2} provides a DFT-based estimate of the energy required to break the Si-C anchoring. That study sets one contribution to $E_1$ as we also assume four Si anchors per 1 nm length of the patch, consistent with  the motif study (Fig. \ref{SI_fig_DFT_2_stick_and_ball}). The impact of this anchoring is clearly evident for narrow-width cases (see Fig. \ref{SI_fig_DFT_energy_diagram1}). Still, the competing $E_2$ term quickly overpowers that $E_1$ binding contribution as the terrace width increases.

The expected crossover in exfoliation behavior, Figure \ref{SI_fig_DFT_potential}, is nevertheless delayed to relatively large terrace widths because there is a second contribution to $E_1$: The vdW adhesion between the buffer and the patch. The motif that we have identified by a DFT simulation, Figure \ref{SI_fig_DFT_2_stick_and_ball}, however, has  too little representation of an actual buffer that we can compute the strength of that second vdW adhesion. Instead we note that buffer atoms are involved in strong  bindings with substrate Si atoms and have less electron density
protruding in the region between the buffer and patch. This suggests a slightly weaker vdW attraction  \cite{Hyldgaard2020ScreeningFoundation} between buffer and patch. For the illustration in Fig. \ref{SI_fig_DFT_potential}, we have set this reduction in the patch/buffer vdW adhesion (that contributes to $E_1$) to 95\% of the $E_2$ strength.

Finally, we repeat that the expected outcome, namely patch-free exfoliation, has an exponential sensitivity to the accuracy with which we can assert our total-system-energy form. It is not realistic to seek a quantitative determination of the width that causes a crossover in behavior. Instead, we present our model and illustrations as a framework to interpret the role of various interaction mechanisms.



\newpage


\begin{figure} 
	\centering
	\includegraphics{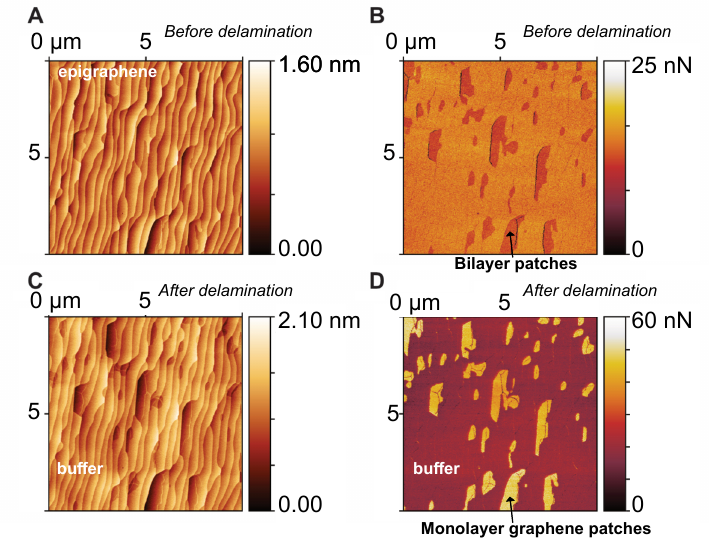} 
	\caption{\textbf{Atomic Force microscopy characterization of graphene on SiC before and after delamination of the top graphene layer, demonstrating that all graphene patches remain on the substrate after graphene delamination.}  (\textbf{A}) Topography map, scan size $10\times10$ $\mu$m$^2$, of graphene on SiC. (\textbf{B}) Adhesion force mapping on graphene on SiC, measured separately along with the topography measurement in A. Dark color represents bilayer patches on graphene. (\textbf{C}) Topography scan of the same area as in A and B but after delamination of the top graphene layer. (\textbf{D}) Adhesion map of graphene on SiC, measured along with topography in C. The lighter color corresponds to the remaining (100\%) monolayer graphene patches.
    }
	\label{subfig:1} 
\end{figure}

\begin{figure} 
	\centering
	\includegraphics{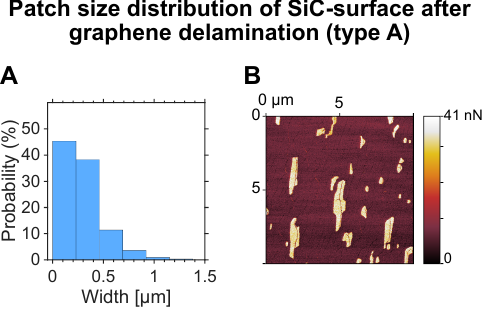} 
	\caption{\textbf{Histograms obtained from AFM data showing the sizes of the remaining patches after peeling the top graphene layer from a Type A-epigraphene surface. The data (300 counts) are normalized to probability.}
   (\textbf{A}) Distribution of patch widths. (\textbf{B}) An AFM adhesion map representative for the data set used to obtain the histograms.}
	\label{SI_fig_histogram_patchsize_typeA} 
\end{figure}

\begin{figure} 
	\centering
	\includegraphics{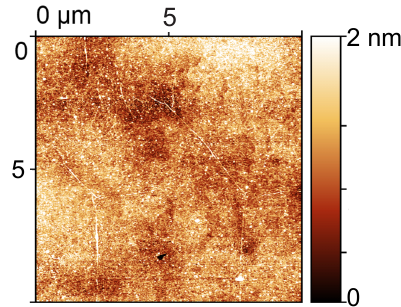} 
	\caption{\textbf{AFM height map of graphene transferred onto SiO$_2$/Si.} There is presence of small wrinkles and adsorbates, likely polymeric and metallic contaminants. We also note that the transferred graphene layer does not inherit the stepped morphology of as-grown epigraphene on SiC.
    }
	\label{supplementary_fig:AFM_transferred} 
\end{figure}

\begin{figure} 
	\centering
	\includegraphics{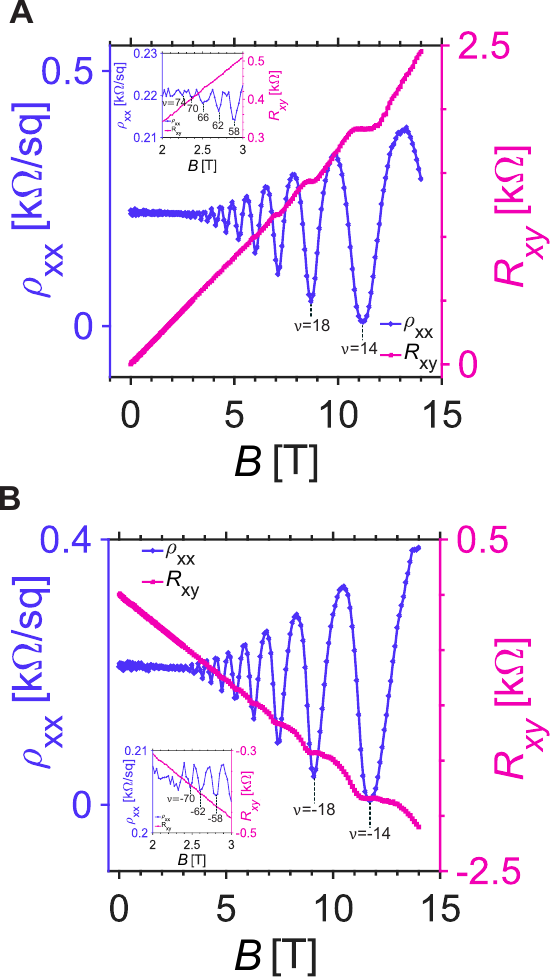} 
	\caption{\textbf{{Shubnikov-de Haas-oscillations and quantum Hall plateaus measured at 2 K for the device shown in Fig. \ref{fig:3}A.}} (\textbf{A}) Electrons  ($V_g = 60$ V) at carrier densities $n_H =3.73 \times 10^{12}$ cm$^{-2}$ from Hall slope and $n_s =3.77 \times 10^{12}$ cm$^{-2}$ from Eq. \ref{ns}. The inset shows the smallest oscillations (highest $N$, up to $N = 74$).
    (\textbf{B}) Holes ($V_g = -44$ V) at $n_H =4.02 \times 10^{12}$ cm$^{-2}$ and $n_s =3.92 \times 10^{12}$, with the inset showing the smallest oscillations, up to $N=70$.}
	\label{subfig:sdH_small_device_holes} 
\end{figure}

\begin{figure} 
	\centering
	\includegraphics[width=0.45\textwidth]{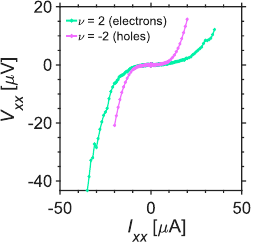} 
	\caption{\textbf{IV-curve measured at $B=14$ T at the $\nu = \pm 2$-QH plateaus for holes ($V_g = 15$ V) and $V_g = 35$ V (electrons) for the large device (Fig. \ref{fig:3}G).} Constant DC offsets (mean value of voltage in the flat region $-3\,\mu\mathrm{A} < I_{xx} < 3\,\mu\mathrm{A}$) were subtracted (0.044 $\mu$V for electrons and 0.959 $\mu$V for holes).} 
	\label{subfig:IV} 
\end{figure}

\begin{figure} 
	\centering
	\includegraphics{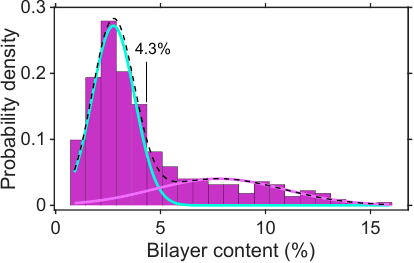} 
	\caption{{\textbf{Histogram of the bilayer content on the 4'' graphene wafer transferred from SiC to SiO$_2$/Si from the mapping in Fig. \ref{fig:4}E.} The counts are normalized to probability density and fitted to a bimodal model consisting of two Gaussian distributions. The median of all values is 4.3$\%$. For the first fit, $\mu_1 = 2.8\%$, $\sigma_1 =1.0\%$ and for the second, $\mu_2 = 7.8\%$, $\sigma_2 =3.1\%$. The dashed line represents the composite fit.}
  }
	\label{SI_fig_histogram_Bilayer_content_wafer_mapping} 
\end{figure}

\begin{figure} 
	\centering
	\includegraphics{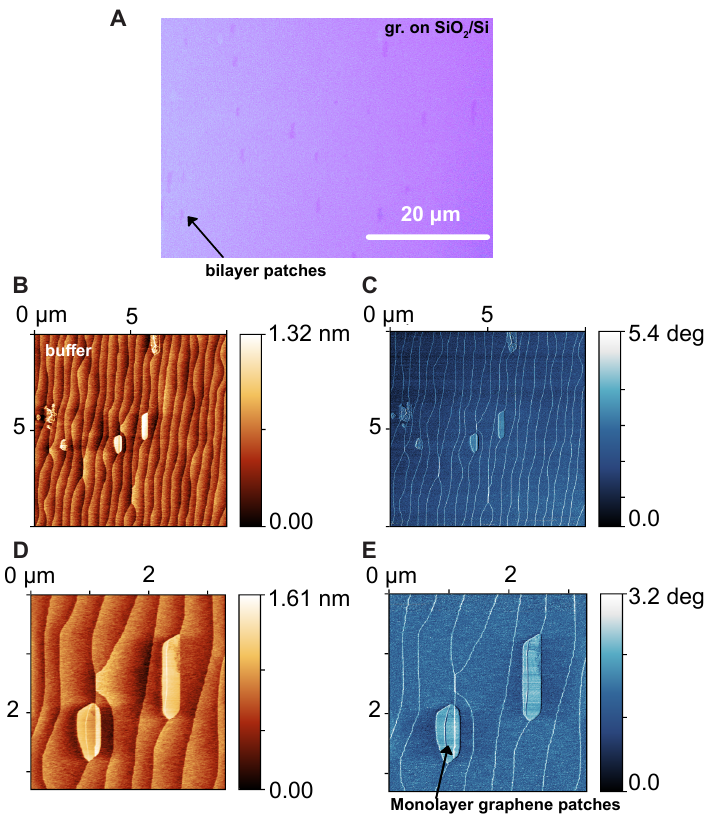} 
	\caption{\textbf{AFM characterization of the SiC substrate demonstrates that small (up to micron-sized) graphene patches remain on the substrate after nickel-assisted graphene delamination from a Type A surface.} (\textbf{A}) Optical micrograph of graphene transferred from SiC onto SiO$_2$/Si following delamination from a Type A SiC substrate using 200 nm nickel. (\textbf{B}) AFM topography scan, size $10\times10$ $\mu$m$^2$, of the SiC surface after delaminating the graphene layer using nickel. (\textbf{C}) Phase contrast mapping of C where light blue represents monolayer patches remaining after delamination. (\textbf{D}) A smaller-area AFM height map of the area in B-C. (\textbf{E}) Phase map of D), revealing patches with wrinkles that indicate that patches were about to be delaminated, but stayed on the substrate. 
    }
	\label{SI:AFM_Ni} 
\end{figure}

\clearpage

\begin{table}
    \centering
    \caption{\textbf{Fitted FET mobilities from Fig.\ref{fig:3}B and impurity densities estimated from these using $\mathbf{\mu_{FET} = 50\,( n_0 / n_{\mathrm{imp}})\mu_0}$ \cite{Adam2007ATransport} (assuming electron/hole symmetry).} }
    \label{SI_tab:imp_density_mobility_FET}

     \begin{tabular}{l cc}
        \hline
        Temperature & $\mu_{\mathrm{FET}}$ [cm$^{2}$/Vs] & $n_{\mathrm{imp}}$ [cm$^{-2}$] \\
        \hline
        300 K & 8{,}670  & $5.77 \times 10^{11}$ \\
        2 K   & 11{,}000 & $5.10 \times 10^{11}$ \\
        \hline
    \end{tabular}
\end{table}

\begin{table}
    \centering
    \caption{\textbf{Fit parameters from fit of Hall mobility vs. carrier density to Eq.\ref{mobility_vs_n_linearized} (see Fig. \ref{fig:3}C), with impurity density from $\mathbf{C = 50\,( n_0 / n_{\mathrm{imp}})\mu_0}$ \cite{Adam2007ATransport}}.}
    \label{SI_table:Hall_fit}

    \begin{tabular}{lccc} 
        \hline
        & C [cm$^{-2}$V$^{-1}$s$^{-1}$] & A/B [cm$^{2}$] & $n_{\mathrm{imp}}$ [cm$^{-2}$] \\
        \hline
        Holes (300 K)    & 8650  & $6.81 \times 10^{-14}$  & $5.78 \times 10^{11}$ \\
        Electrons (300 K) & 8650  & $1.081 \times 10^{-13}$ & $5.78 \times 10^{11}$ \\
        Holes (2 K)      & 9720  & $6.03 \times 10^{-14}$  & $5.15 \times 10^{11}$ \\
        Electrons (2 K)  & 9550  & $5.27 \times 10^{-14}$  & $5.23 \times 10^{11}$ \\
        \hline
    \end{tabular}
\end{table}


\clearpage 





\end{document}